\documentclass[aps,prd,preprint,groupedaddress,showpacs,nofootinbib]{revtex4}

\usepackage{graphicx}
\usepackage{amsmath}

\def\ltwid{\mathrel{\raise.3ex\hbox{$<$\kern-.75em\lower1ex\hbox{$\sim$}}}}
\def\gtwid{\mathrel{\raise.3ex\hbox{$>$\kern-.75em\lower1ex\hbox{$\sim$}}}}

\def\square{\kern1pt\vbox{\hrule height 1.2pt\hbox{\vrule width 1.2pt\hskip 3pt
   \vbox{\vskip 6pt}\hskip 3pt\vrule width 0.6pt}\hrule height 0.6pt}\kern1pt}
\def\overleftrightarrow#1{\vbox{\ialign{##\crcr
     $\leftrightarrow$\crcr\noalign{\kern-1pt\nointerlineskip}
     $\hfil\displaystyle{#1}\hfil$\crcr}}}

\newcommand{\be}{\begin{equation}}
\newcommand{\ee}{\end{equation}}
\newcommand{\bea}{\begin{eqnarray}}
\newcommand{\eea}{\end{eqnarray}}

\begin{document}

\title{Late time solution for self-interacting scalar field in accelerating space-times}
\title{Late time solution for interacting scalar in accelerating spaces}

\author{Tomislav~Prokopec}
\email[]{t.prokopec@uu.nl}

\affiliation{Institute for Theoretical Physics, Spinoza Institute and EMME$\Phi$, Utrecht University,\\
Postbus 80.195, 3508 TD Utrecht, The Netherlands}



\begin{abstract}


We consider stochastic inflation in an interacting scalar field in spatially homogeneous accelerating space-times
with a constant principal slow roll parameter $\epsilon$. We show that, if the scalar potential is scale invariant 
(which is the case when scalar contains quartic self-interaction and couples non-minimally to gravity),
the late-time solution on accelerating FLRW spaces can be described by a probability distribution function (PDF) $\rho$ which is a function 
of $\varphi/H$ only, where 
$\varphi=\varphi(\vec x)$ is the scalar field and $H=H(t)$ denotes the Hubble parameter. 
We give explicit late-time
solutions for $\rho\rightarrow \rho_\infty(\varphi/H)$, and thereby find the order $\epsilon$ corrections to the Starobinsky-Yokoyama 
result. This PDF can then be used to calculate {\it e.g.} various $n-$point functions of the (self-interacting) scalar field,
which are valid at late times in arbitrary accelerating space-times with $\epsilon=$ constant.

\end{abstract}


\maketitle


\section{Introduction}
\label{Introduction}

In 1986 Alexei Starobinsky~\cite{Starobinsky:1986fx} realised that a classical stochastic theory can be used to calculate
infrared correlators of interacting scalar fields on De Sitter space. The stochastic formalism was then 
used to calculate the late-time state of a self-interacting scalar field~\cite{Starobinsky:1994bd}.
Almost 20 years later Tsamis and Woodard~\cite{Tsamis:2005hd} proved that stochastic formalism correctly captures the leading 
logarithms on De Sitter background to all orders in perturbation theory, and therefore can be used to efficiently resum
the perturbation theory on De Sitter in scalar interacting theories. 
Miao and Woodard~\cite{Miao:2006pn} and Prokopec, Tsamis and Woodard~\cite{Prokopec:2007ak} have shown how to generalize
the results of~\cite{Tsamis:2005hd}
to include light fermionic and (massless) Abelian gauge fields, respectively. Since massless fermions and Abelian gauge fields 
couple conformally to gravity, they cannot be treated as classical.  The authors of~\cite{Miao:2006pn} and~\cite{Prokopec:2007ak}
realized that these {\it passive} fields (light fermions and Abelian gauge fields) 
can be integrated out to obtain an effective interacting scalar theory,
which can be then stochasticized by the usual Starobinsky's procedure. Explicit two-loop calculations
in scalar electrodynamics~\cite{Prokopec:2008gw,Prokopec:2006ue}
vindicate the stochastic inflationary predictions of~\cite{Prokopec:2007ak}. 
Despite of some notable attempts~\cite{Vennin:2015hra,Finelli:2010sh,Finelli:2008zg,Tsamis:2005hd}, 
it is fair to say that at this moment it is not known how to rigorously stochastize (perturbative) quantum gravity on De Sitter or on 
inflationary space-times. 

  Recently Gihyuk Cho, Cook Hyun Kim and Hiroyuki Kitamoto~\cite{Kitamoto:2015,ChoKimKitamoto:2015} 
showed that the correct way 
to stochasticize light interacting scalar fields on accelerating spaces 
(defined by $0\leq \epsilon=-\dot H/H^2<1$, where $H$ is the Hubble parameter), 
introduces an important order $\epsilon$-correction to the friction in the na\^ive stochastic equations.
In this letter we make use of Cho-Kim-Kitamoto's result to solve for the late-time solution of the field.

We show that scale invariant interacting scalar theories admit at late times 
a solution in which the PDF $\rho=\rho(\varphi/H)\equiv \exp[-v_{\rm eff}(\varphi/H)]$, where 
$v_{\rm eff}(\varphi/H)$ generalises the Starobinsky-Yokoyama solution~\cite{Starobinsky:1994bd}.
Recall that the Starobinsky-Yokoyama solution is of the form, $\rho_{\rm SY}=\rho_0\exp(-v)$,
with $v=[8\pi^2/(3H^4)]V(\varphi)$, where $V(\varphi)=\lambda\varphi^4/4!$ is the tree-level potential. 
Strictly speaking this solution applies to De Sitter space only. In this letter we show 
that in accelerating spaces with a constant $\epsilon$ parameter,  the late-time solution of 
Starobinsky and Yokoyama $\rho_{\rm SY}$ gets modified. Furthermore, we explicitly calculate 
the $\epsilon$-induced corrections to the PDF.
Next, we show that only some classes of potentials
-- namely those that exhibit scale invariance -- permit simple 
late-time solution in the form $\rho\rightarrow \rho_\infty(\varphi/H)$.
Curiously, these are precisely the potentials that coincide with those studied in Ref.~\cite{Janssen:2009pb}
 that admit classical solutions of the form,
$\phi(t) =\psi_0 H(t)$, where $\psi_0$ is some (calculable) constant. In Ref.~\cite{Janssen:2009pb} the one-loop-effective potential 
was calculated for a background field of the form, $\phi(t) =\psi_0 H(t)$. The conclusion of Ref.~\cite{Janssen:2009pb} was that two types of  
quantum effects break the classical scale invariance: (1)  the effects induced by the counterterms in the process of renormalization and 
(2) the effects resulting from regulating the infrared sector of the scalar field (whose na\^ive Bunch-Davies vacuum contains power low divergences 
in the infrared). 
Remarkably, stochastic analysis presented in this letter shows that, when 
perturbative contributions are resummed, {\it the scaling symmetry of the classical theory gets restored},
constituting the main result of this letter.  


\section{The Model and its Stochastic Analysis}
\label{The Model and its Stochastic Analysis}

 Consider a scalar field $\phi$ whose dynamics is implied by the (classical) action,
\begin{eqnarray}
 S[\phi] = \int d^4 x\sqrt{-g}\left[-\frac{1}{2}g^{\mu\nu}\partial_\mu\phi\partial_\nu\phi -V(\phi)-\frac12\xi R\phi^2\right]
\,,\quad
\label{classical action}
\end{eqnarray}
where $g_{\mu\nu}$ is the metric tensor, $g^{\mu\nu}$ its inverse, the metric signature is ${\rm sign}[g_{\mu\nu}]=(-1,1,1,1)$,
$R=R(g_{\mu\nu})$ is the Ricci scalar, and $V(\phi)$ is
the (tree level) scalar potential. When gravity is taken to be non-dynamical and when the scalar field is assumed to be light,
 $ |[\partial^2 V/\partial\phi^2]_{\phi=0}|\ll H^2$, $|\xi| \ll 1/12$,  the scalar theory~(\ref{classical action})
on De Sitter space
and in the infrared (on super-Hubble scales) is well approximated by 
Starobinsky's stochastic inflation~\cite{Starobinsky:1986fx,Starobinsky:1994bd}.
In this letter we study the late-time dynamics of the scalar field by making use of the modified 
stochastic inflation first obtained in~\cite{Kitamoto:2015,ChoKimKitamoto:2015}.
To make it suitable for stochastic treatment, the coupling constants $\xi$ and the corresponding couplings in the potential $V(\phi)$ 
are the renormalized couplings evaluated at the physical momentum scale $\bar\mu \simeq \mu H_0$, where $\mu\ll 1$ and $H_0$ is the 
Hubble parameter at a suitably chosen initial (sufficiently early) time. 

Here we work in a flat FLRW space-time, for which the metric tensor can be written as,
\begin{equation}
 g_{\mu\nu} = -dt^2 + a^2(t) d\vec x\cdot d\vec x
\,,
\label{metric}
\end{equation}
where $a=a(t)$ is the scale factor. In terms of the scale factor, one can define the Hubble parameter, $H(t) =\dot a/a$
and the principal slow roll parameter, $\epsilon = -\dot H/H^2$. The Ricci scalar is then,
$R=6(2-\epsilon)H^2$. In De Sitter space, $\dot H=0$, such that $\epsilon=0$ and $R=12H^2$ is constant.

\subsection{Stochastic theory on De Sitter space}
\label{Stochastic theory on De Sitter space}

Stochastic inflation on De Sitter can be obtained by the following simple procedure (which is not a derivation;
for a derivation see~\cite{Tsamis:2005hd,Prokopec:2007ak}).
The action~(\ref{classical action}) implies the following operator equation of motion for $\phi$,
\begin{equation}
  \ddot \phi(t,\vec x) + 3H\dot \phi -\frac{\nabla^2}{a^2}\phi +V'(\phi)+\xi R\phi = 0
\,,
\label{operator EOM phi}
\end{equation}
where $V' = dV/d\phi$.
Here we are interested in the dynamics of the infrared part $\Phi(t,\vec x)$
of the field $\phi$. In the limit when the scalar field is light, $|m^2_\phi|=|d^2V/d\phi^2(\phi=0)+\xi R\ll H^2|$,
the scalar field dynamics is correctly described by the following simple stochastic equation of the Langevin type,
\begin{equation}
 \dot \Phi +\frac{V'( \Phi)-V''(0)\Phi}{3H} = \frac{1}{3H}\dot\Phi_0
 \,,
\label{stochastic equation:dS}
\end{equation}
where the couplings in $V(\Phi)$ and $\xi$ are to be evaluated at the ultraviolet cutoff scale of the theory, given by $\tilde \mu=\mu H_0$
($\mu\ll 1$), and
$\Phi_0(t,\vec x)$ is the solution of the homogeneous linear equation,
\begin{equation}
  \ddot \Phi_0(t,\vec x) + 3H\dot \Phi_0 -\frac{\nabla^2}{a^2}\Phi_0
  +\big[6\xi(2\!-\!\epsilon)H^2+V''(0)\big]\Phi_0
= 0
\,.
\label{operator EOM phi0}
\end{equation}
This linear equation can be solved by decomposing the field in terms of  the mode functions $u(t,k)$ and $u^*(t,k)$ as,
\begin{equation}
\Phi_0(t,\vec x) = \int \frac{d^3k}{(2\pi)^3}\Theta(\mu Ha-k)\left[u(t,k)\hat a(\vec k) {\rm e}^{i\vec k\cdot\vec x}
  + u^*(t,k)\hat a^\dagger(\vec k) {\rm e}^{-i\vec k\cdot\vec x} \right]
  \,,
\label{Phi 0}
\end{equation}
where $\hat a(\vec k)$ and $\hat a^\dagger(\vec k)$ are the annihilation and creation operators
and $\mu\ll 1$ is the ultraviolet cutoff scale. The operator
$\hat a(\vec k)$  annihilates the free vacuum state $|\Omega\rangle$ of the theory, $\hat a(\vec k)|\Omega\rangle=0$, and
$\hat a^\dagger(\vec k)$ creates a particle of momentum $\vec k$ out of the vacuum. Applying
successively  $\hat a^\dagger(\vec k)$ on $|\Omega\rangle$ generates the Fock space of the theory.

 On De Sitter space and in the massless limit ($\epsilon=0$, $\xi R+V'' (0)=0$), the properly normalized mode functions
 (which obey the canonical Wronskian, $W[u,u^*]=i/a^3$), are given by,
\begin{equation}
 u(t,k) = \frac{H}{\sqrt{2k^3}}\left[1\!-\!\frac{ik}{aH}\right]\exp\Big(i\frac{k}{aH}\Big)
 \,,\quad
 u^*(t,k) = \frac{H}{\sqrt{2k^3}}\left[1\!+\!\frac{ik}{aH}\right]\exp\Big(\!-i\frac{k}{aH}\Big)
 \,.
\label{mode function:dS}
\end{equation}
%
In the more complicated case, when the scalar is massive, the mode functions can be expressed in terms of the Hankel functions
of the first and second kind. From the solutions~(\ref{mode function:dS}) it is clear that the first term
in the square brackets dominate in the infrared (on super-Hubble scales), when $k/a\ll H$, such that
the mode functions in $\Phi_0$ in Eq.~(\ref{Phi 0}) can be approximated by, $u(t,k)=u^*(t,k)\simeq H/\sqrt{2k^3}$.
With this approximation in mind, one can show that the field operator $\Phi_0$ becomes a classical stochastic
field, in the sense that its canonical commutator vanishes, $[\Phi_0(t,\vec x),\dot\Phi_0(t,\vec y)]=0$,
and its correlator is Markovian,
\begin{equation}
  \langle\Omega|\xi(t,\vec x)\xi(t',\vec y)|\Omega\rangle
     =\frac{(\mu a)^2(\mu Ha\dot{)} }{18\pi^2}\frac{\sin[\mu H a r]}{\mu H ar}|u(t,k=\mu H a)|^2\delta(t-t')
     \,,\quad r =\|\vec x\!-\!\vec y\|
     \,,
\label{stochastic force}
\end{equation}
where
$\xi(t,\vec x)=\dot\Phi_0/(3H)$. The formula~(\ref{stochastic force}) simplifies further when one takes
account of the infrared approximation for the mode functions,
$|u(t,k)|^2\rightarrow 1/[2H (\mu a)^3]$ and in the limit of spatial coincidence,
when $\sin[\mu H a r]/(\mu H ar)\rightarrow 1$. In this case the noise correlator~(\ref{stochastic force}) simplifies to,
$\langle\Omega|\xi(t,\vec x)\xi(t',\vec y)|\Omega\rangle \simeq [H/(36\pi^2)]\delta(t-t')$.

\subsection{Stochastic theory on accelerating spaces}
\label{Stochastic theory on  accelerating spaces}

Cho, Kim and Kitamoto recently showed~\cite{Kitamoto:2015,ChoKimKitamoto:2015} that, in order to obtain the stochastic theory
applicable to accelerating spaces with constant $\epsilon$ (for which
 $0\leq \epsilon<1$), it is convenient to rewrite the operator equation~(\ref{operator EOM phi})  as
\begin{equation}
  \partial_N^2\phi(t,\vec x) + (3-\epsilon)\partial_N\phi -\frac{\nabla^2}{(aH)^2}\phi +\frac{V'(\phi)+\xi R\phi}{H^2} = 0
\,,
\label{operator EOM phi: N}
\end{equation}
where $N=\ln(a)$ is the number of e-foldings. The corresponding stochastic theory is then obtained by
dropping the second derivative terms in~(\ref{operator EOM phi: N}) and adding a stochastic force $\xi$ as,
\footnote{There is no unique way to split the potential into the one that gives rise to the stochastic force 
and the interactions. In Eq.~(\ref{stochastic equation:accelerated}) we use the standard procedure, 
and include the quadratic terms in the action into the equation of motion for the free field $\Phi_0$, and all cubic and higher order 
terms into the interactions. This split is supported by the well known fact that large masses suppress stochastic force, hence it is natural 
to include them into $\Phi_0$. In the Appendix we consider another possibility, according to which 
mass terms are split between the interactions and the free field. That choice is motivated 
by the desire to retain the cutoff independence of the De Sitter case.}
\begin{equation}
 \partial_N\Phi(N,\vec x)
 +\frac{V'(\Phi)-V''(0)\Phi
}{(3-\epsilon)H^2} = \frac{\dot\Phi_0}{(3-\epsilon)H}\equiv \xi(t,\vec x)
 \,,
 \label{stochastic equation:accelerated}
 \end{equation}
where the stochastic force $\xi$ obeys,
\begin{equation}
  \langle\Omega|\xi(t,\vec x)\xi(t',\vec y)|\Omega\rangle
     =\frac{(\mu a)^3H^2(1-\epsilon) }{(3-\epsilon)^2 2\pi^2}\frac{\sin[\mu H a r]}{\mu H ar}|u(t,k)|^2\delta(t-t')
     \,,
\label{stochastic force:accelerated}
\end{equation}
and $u$ is the mode function satisfying,
\begin{equation}
  \Big(\partial_\eta^2+k^2-\frac{a''}{a}+a^2\big[V''(0)+6\xi(2-\epsilon)H^2\big]\Big)[au(\eta,k)]=0.
\label{mode equation:accelerated}
\end{equation}
For a constant epsilon, $a''/a=(2\!-\!\epsilon)/[\eta^2(1-\epsilon)^2]$
and assuming $m_\phi/H={\rm const.}$ (we will see below in which situations
this approximation holds), we get
\begin{equation}
  \Big(\partial_\eta^2+k^2-\frac{(2\!-\!\epsilon)(1\!-\!6\xi)}{(1\!-\!\epsilon)^2}\frac{1}{\eta^2}
  +a^2V''(0)\Big)[au(\eta,k)]=0.
\label{mode equation:accelerated:2}
\end{equation}
The Bunch-Davies (or Chernikov-Tagirov) vacuum solution~\footnote{The mode functions for
the most general homogeneous pure Gaussian state can be written as, 
$\alpha(k)u(\eta,k)+\beta(k)u^*(\eta,k)$, with $|\alpha|^2-|\beta|^2=1$.}
 for the mode functions is (see {\it e.g.} Refs.~\cite{Janssen:2009pb}  and~\cite{Glavan:2014uga}),
\begin{eqnarray}
 u(\eta,k) &=& \frac1a\sqrt{\frac{\pi}{4(1\!-\!\epsilon){\cal H}}}H_\nu^{(1)}\Big(\frac{k}{(1\!-\!\epsilon){\cal H}}\Big)
 \nonumber\\
 u^*(\eta,k)&=&\frac1a\sqrt{\frac{\pi}{4(1\!-\!\epsilon){\cal H}}}H_\nu^{(2)}\Big(\frac{k}{(1\!-\!\epsilon){\cal H}}\Big)
\,,\qquad \nu^2 =\frac14+\frac{(2-\epsilon)(1-6\xi)}{(1-\epsilon)^2}-\frac{V''(0)}{H^2}
\,,
\label{mode functions: cons eps}
\end{eqnarray}
where ${\cal H}=d[\ln(a(\eta))]/d\eta=aH$ is the conformal Hubble rate.
These solutions are correct provided $V''/H^2$ are time independent (in slow roll inflation one allows for a
weak dependence of $\nu$ on time, in the sense that $|\dot \nu/\nu|\ll H$, which amounts to an adiabatic approximation).
Since mode functions in~(\ref{stochastic force:accelerated}) are truncated at $k/a=\mu H\ll H$, one can approximate
the Hankel functions in~(\ref{mode functions: cons eps}) with their small argument expansion.~\footnote{The relevant small argument 
expansion of the Hankel functions is, 
$H_\nu^{(1)}(z)=[H_\nu^{(2)}(z)]^*\simeq [\Gamma(\nu)^2/\pi](2/z)^{2\nu}+{\cal O}(z^{-2\nu+2},z^0,z^{2\nu})$ 
($\Im[\nu]=0$, $\nu>0$, $\Im[z]=0$).
}
When the leading term in this expansion is inserted into~(\ref{stochastic force:accelerated})
one obtains,
\begin{equation}
  \langle\Omega|\xi(t,\vec x)\xi(t',\vec y)|\Omega\rangle=\frac{ H^2A(\epsilon,\nu,\mu)}{(3\!-\!\epsilon)^2}
                   \frac{\sin[\mu H a r]}{\mu H ar}\delta(N(t)-N(t'))
  \,,
\label{stochastic force:accelerated:2}
\end{equation}
where
\begin{equation}
  A =\frac{ \mu^{3-2\nu}(1\!-\!\epsilon)^{2\nu}2^{2\nu-3}\Gamma(\nu)^2}{\pi^3}
    \,,
\label{stochastic force:accelerated:3}
\end{equation}
such that, when $\nu\neq 3/2$, stochastic force becomes dependent on the cutoff scale $\mu$. If we are interested in the cases
when $|\epsilon|\ll 1$, $12|\xi|\ll 1$ and $|V''(0)|\ll H^2$, then $\nu\approx 3/2$ ($|\nu-3/2|\ll 1$), the dependence on the UV cutoff is 
weak, and stochastic treatment applies. One could argue that the better choice for the UV cutoff in Eq.~(\ref{Phi 0}) is 
$k_{\rm UV}=\mu (1-\epsilon)aH$. Indeed, for this choice the argument of the Hankel function becomes $\mu$, which equals
to $1$ when $\mu=1$, signifying the scale at which mode functions start coupling strongly to the background, and the process of large 
(non-adiabatic) particle creation sets in. For that choice of the cutoff, $A$ in Eq.~(\ref{stochastic force:accelerated:3}) would 
rescale as, $A\rightarrow A(1\!-\!\epsilon)^{3-2\nu}$, retaining thus the same dependence on the cutoff scale $\mu$,
thus not changing~(\ref{stochastic force:accelerated:3}) in any significant way. 

Just like in De Sitter space, the stochastic theory~(\ref{stochastic equation:accelerated}--\ref{stochastic force:accelerated})
can be recast as a Fokker-Planck equation~\cite{Starobinsky:1994bd}
for the probability distribution function (PDF) $\rho=\rho(N(t),\varphi(\vec x))$ (for the stochastic field $\varphi(\vec x)$
which represents the quantum field $\Phi(t,\vec x)$),
\begin{equation}
 \partial_N \rho(N,\varphi(\vec x)) = \frac{H^2A}{2}\partial_\varphi^2\rho+\frac{1}{(3\!-\!\epsilon)H^2}\partial_\varphi[\rho V'(\varphi)]
\,,
\label{Fokker-Planck equation}
\end{equation}
where we made use of $\partial_t=H\partial_N$, which follows from $N=\ln(a)$.
Starobinsky and Yokoyama~\cite{Starobinsky:1994bd} 
told us how to solve the Fokker-Planck equation in the case of a self-interacting massless scalar field on De Sitter space,
for which $V(\phi)=[\lambda/4!]\phi^4$. Their solution can be easily extended to the case of a light scalar. even though in that
limit the solution picks up a slight dependence on the cutoff scale $\mu$.

\subsection{Stochastic late-time solution}
\label{Stochastic late-time solution}

In order to study asymptotically late-time solutions of~(\ref{Fokker-Planck equation})
in an accelerating space-time with constant epsilon,
it is convenient to transform from $(N,\varphi)$ to the variables $(N,\tilde\varphi=\varphi/H)$,
in which case,
\begin{equation}
\partial_N \rightarrow \partial_N + \epsilon\tilde\varphi\partial_{\tilde\varphi}
\,.\qquad
\partial_\varphi\rightarrow H^{-1}\partial_{\tilde\varphi}
\,.
\label{Fokker-Planck equation:coord transform}
\end{equation}
Furthermore, from $\int d\varphi \rho(N,\varphi) =  \int d\tilde\varphi \tilde\rho(N,\tilde\varphi)$,
it follows that $\tilde\rho=H\rho$.~\footnote{I thank Hiroyuki Kitamoto 
for pointing out to me that this rescaling of $\rho$ significantly simplifies the late-time solution for the field PDF.}
With these in mind, Eq.~(\ref{Fokker-Planck equation}) becomes,
\begin{equation}
  \partial_N\tilde\rho(N,\tilde\varphi) +\epsilon\partial_{\tilde\varphi}\big[\tilde\varphi\tilde\rho(N,\tilde\varphi)\big]
   =   \frac{A}{2}\partial_{\tilde\varphi}^2\tilde\rho+\frac{1}{3\!-\!\epsilon}\partial_{\tilde\varphi}[\tilde\rho\tilde V'(\tilde\varphi)]
   \,,
\label{Fokker-Planck equation:2}
\end{equation}
where $\tilde V(x)$ is defined by,
\begin{equation}
\tilde V(\tilde\varphi) \equiv \frac{V(H\tilde\varphi)}{H^4}
\,.
\label{rescaled potential}
\end{equation}
A closer inspection of~(\ref{Fokker-Planck equation:2}) implies that, if we are to have a time independent solution at late times,
then time dependence  must disappear from the right hand side of~(\ref{Fokker-Planck equation:2}), {\it i.e.} $\tilde V(\tilde\varphi)$
must be time-independent. This is the case when the scalar field is invariant under scale transformations,~\footnote{
By the invariance under scale transformations we mean the invariance of the classical action under the transformations of the type,
$g_{\mu\nu}\rightarrow \Omega^2 g_{\mu\nu}$, $\varphi\rightarrow \Omega^{-1}\varphi$, where $\Omega$ is an arbitrary constant.
It then follows that the Ricci scalar transforms as, $R\rightarrow \Omega^{-2} R$, $\sqrt{-g}\rightarrow \Omega^4\sqrt{-g}$
and the action~(\ref{classical action}) is invariant
if $V\rightarrow \Omega^{-4} V$, which is the case when $V=\lambda\varphi^4/4!$.
}
 {\it i.e.} when $V(\phi)=[\lambda/4!]\phi^4$.
In this case,
\begin{equation}
\tilde V(\tilde\varphi) = \frac{\lambda}{4!}\tilde\varphi^4
\,.
\label{rescaled potential}
\end{equation}
For example, neither cubic interactions, $\lambda_3\varphi^3/3!$, nor tree mass terms, $m^2\varphi^2/2$,
 are not allowed in the potential, because it would generate the terms,
 $[m/H(t)]^2\tilde\varphi^2/2+(\lambda_3/6H)\tilde\varphi^3$, in $\tilde V$, which depend on time through $H(t)$.

With these comments in mind, the late-time limit of Eq.~(\ref{Fokker-Planck equation:2}) becomes,
\begin{equation}
 \partial_{\tilde\varphi}[\epsilon\tilde\varphi\tilde\rho_\infty(\tilde\varphi)]
   =   \frac{A}{2}\partial_{\tilde\varphi}^2\tilde\rho_\infty
   +\frac{1}{3\!-\!\epsilon}\partial_{\tilde\varphi}
                    \Big[\tilde\rho_\infty(\tilde\varphi)\Big(\frac{\lambda}{6}\tilde\varphi^3\Big)\Big]
   \,.
\label{Fokker-Planck equation:2bb}
\end{equation}
This equation governs the late-time PDF of the scalar field in stochastic inflation.
Next, it is convenient to rewrite Eq.~(\ref{Fokker-Planck equation:2bb}) in the form,
\begin{equation}
  \partial_{\tilde\varphi}^2\tilde\rho_\infty + \partial_{\tilde\varphi}\big[ v_{\rm eff}' \tilde\rho_\infty\big]=0
\,,\qquad v_{\rm eff} = \frac{2}{(3\!-\!\epsilon)A}\left[\tilde V(\tilde\varphi)-\frac{(3\!-\!\epsilon)\epsilon}{2}\tilde\varphi^2\right]
\,.
\label{Fokker-Planck equation:2b}
\end{equation}
Then the following transformation, $\tilde\rho_\infty = \tilde\rho_0\exp[-v(\tilde\varphi)]$ yields the differential equation for $v(\tilde\varphi)$,
\begin{equation}
  -v''+v_{\rm eff}''+[v'-v_{\rm eff}']v' = 0
\,.
\label{Fokker-Planck equation:2c}
\end{equation}
The obvious (and correct) solution is, 
\begin{equation}
  v(\tilde\varphi)=v_{\rm eff}(\tilde\varphi)
\,,
\label{Fokker-Planck equation:2d}
\end{equation}
where $v_{\rm eff}(\tilde\varphi)$ is given in~Eq.~(\ref{Fokker-Planck equation:2b}).
In conclusion, the evolution of an interacting scalar field in accelerating spaces gets modified such that -- due to the deviation from De Sitter 
exponential expansion -- at late times the field acquires a negative, time dependent, mass-squared, 
\begin{equation}
 \Delta m^2_\epsilon = -(3\!-\!\epsilon)\epsilon H^2
\label{Fokker-Planck equation:2e}
\end{equation}
Since our field already has a (time dependent) mass-squared, $R=6\xi(2\!-\!\epsilon)H^2$, the total mass becomes, 
\begin{equation}
 m^2_\phi\stackrel{t\rightarrow \infty}{\longrightarrow}\; \big[6\xi(2\!-\!\epsilon)-(3\!-\!\epsilon)\epsilon\big] H^2
\,.
\label{Fokker-Planck equation:2f}
\end{equation}
\vskip -0.15cm\noindent
Depending on the sign and magnitude of $\xi$, this mass-squared can be either positive or negative, indicating that 
deviations from De Sitter can be responsible for generation of symmetry breaking.
A complete understanding of this interesting question requires further study (for a discussion 
of symmetry restoration on De Sitter see~Refs.~\cite{Lazzari:2013boa,Serreau:2013eoa,Guilleux:2015pma}).

 The PDF $\rho(\varphi)$ that we obtained in this letter can be used to calculate asymptotically late-time coincident field correlators
in accelerating spaces. In order to obtain the corresponding renormalized correlators, $\langle\phi^{2n}(x)\rangle$, 
the following formula can be used, 
\begin{equation}
  \langle\phi^{2n}(x)\rangle \simeq \langle \varphi^{2n}\rangle_{\rm stoch} =\int d\varphi \rho(\varphi) \varphi^{2n}
 = H^{2n}\int d\tilde\varphi \tilde\rho(\tilde\varphi)\tilde \varphi^{2n}
\,,
\label{correlators}
\end{equation}
whereby $\rho$ is assumed to be suitably normalized, $\int d\varphi \rho(\varphi) =1$. The relevant integrals 
can be expressed in terms of the confluent hypergeometric functions and the modified Bessel functions~\cite{Lazzari:2013boa}.

 \section{Conclusion and Discussion}
 \label{Conclusion and Discussion}
  
\hskip -0.15cm\noindent
  In this letter we have constructed the late-time solution for an interacting scalar field theory on accelerating spaces 
  with a constant principal slow roll parameter, $\epsilon\equiv -\dot H/H^2$. The most general tree level potential
  for which one can construct an exact solution contains quartic self-coupling and a non-minimal coupling to gravity.
Our solution~(\ref{Fokker-Planck equation:2d}) shows that, 
the effective potential that governs the PDF at late times 
acquires a negative, order $\epsilon$ mass-squared, $\Delta m^2_\epsilon = -(3\!-\!\epsilon)\epsilon H^2$, 
see Eq.~(\ref{Fokker-Planck equation:2e}), thus effectively increasing the size of field fluctuations.
If the total mass-squared becomes negative, that would suggest a symmetry breaking induced by the expansion.
It would be of interest to understand the dynamics of such a symmetry breaking. Even in the simpler De Sitter background 
the problem of symmetry breaking/restoration has not fully been understood~\cite{Lazzari:2013boa,Serreau:2013eoa,Guilleux:2015pma}. 
The reason why we were able to find an exact late-time solution for the field PDF
is because the tree-level potential is invariant under scale transformations. The resulting effective
potential~(\ref{Fokker-Planck equation:2d}), (\ref{Fokker-Planck equation:2b})
 inherits the scale invariance of the tree-level. 
 
 An interesting question is how our results get modified when one includes into the tree potential 
terms that break scale invariance, such as a mass term or a cubic self-interaction. 
 It is easily seen that in that case the Fokker-Planck equation~(\ref{Fokker-Planck equation}) 
does not admit late-time scale invariant solutions
 simply because in this case the rescaled potential $\tilde V$~(\ref{rescaled potential}) contains time dependent terms 
 $\tilde V\supset m^2\tilde \varphi^2/[2H(t)^2]\propto a^{2\epsilon}, \lambda_3\tilde \varphi^2/[6H(t)]\propto a^{\epsilon}$. 
It would be of interest to construct adiabatic solutions in 
 that more complicated case. We suspect that is possible only when $\epsilon$ is small, {\it i.e.} $\epsilon\ll 1$. 
For a discussion of that question see Ref.~\cite{ChoKimKitamoto:2015}.
 
 One annoying feature of our solution~(\ref{Fokker-Planck equation:2d}), (\ref{Fokker-Planck equation:2b})
 is the cutoff dependence of our late-time solution for the PDF $\rho(\varphi)$ , which is hidden 
in the $\mu$-dependence of the stochastic force parameter~(\ref{stochastic force:accelerated:3}), 
$A\propto \mu^{3-2\nu}$. That dependence is weak  
when $\nu\simeq 3/2$, which is the case during slow roll inflation and when the field mass is negligibly small.
There is a formal way of getting rid of this $\mu$ dependence, which is discussed in the Appendix. 
The idea is to split the mass term into two parts, such that the stochastic force is equal to that of the massless field in De Sitter limit.
Just like in the De Sitter case, the dependence on $\mu$ then disappears.  The resulting late-time PDF gets modified as discussed in
the Appendix such that the effective potential gains a quadratic term with a well defined potential, 
see Eqs.~(\ref{Appendix:Fokker-Planck equation:2b}) and~(\ref{Appendix:Fokker-Planck equation:2d}). 
While this may be 
just a curiosity, it may contain important physics. Namely, to study properly what happens in the deep infrared on accelerating space-times,
an RG treatment should be appropriate. There have been some attempts to apply RG methods in 
De Sitter space~\cite{Lazzari:2013boa,Serreau:2013eoa,Guilleux:2015pma},
but none so far is attempted in more general FLRW space-times. While early results are promising, 
further developments will show how important these results are and whether these techniques can be used to obtain
results that go beyond stochastic inflation, and thus illuminate the curious results obtained in the Appendix.

 The late-time solutions we have constructed here can be of help to better understand  late-time dynamics of the inflaton 
and other interacting scalar fields during inflation, especially in the regime when $\epsilon$ is not very small.
Our effective potential can be used to calculate, for example, late-time limit of $n-$point functions
of the scalar field, by making use of Eq.~(\ref{correlators}).

 \section*{\bf Acknowledgements}

This work is part of the D-ITP
consortium, a program of the Netherlands Organization for Scientific
Research (NWO) that is funded by the Dutch Ministry of Education,
Culture and Science (OCW). 
 
\appendix
\section*{Appendix}

 Here we provide a formal procedure for construction of a cutoff-independent late-time PDF of an interacting scalar field, whose potential 
is scale invariant, 
\begin{equation}
   V(\varphi) = \frac{6\xi(2\!-\!\epsilon)}{2}H^2\varphi^2 +\frac{\lambda}{4!}\varphi^4
\,.
\label{Appendix: potential}
\end{equation}
The cutoff dependence is generated by the fact that free mode functions 
satisfy an equation for which the index $\nu$ defined in Eq.~(\ref{mode functions: cons eps}) differs from $3/2$.
To get $\nu=3/2$, one can split the mass term in the potential~(\ref{Appendix: potential}) as,
\begin{equation}
   V(\varphi) =(\omega_1+\omega_2) \frac{6\xi(2\!-\!\epsilon)}{2}H^2\varphi^2 +\frac{\lambda}{4!}\varphi^4
\,,\qquad \omega_1+\omega_2 =1 
\nonumber
\end{equation}
such that $\nu=3/2$, or equivalently, 
\begin{equation}
\nu^2 =\frac14+\frac{(2-\epsilon)(1-6\omega_1\xi)}{(1-\epsilon)^2}=\frac94
\nonumber
\end{equation}
 is achieved by 
\begin{equation}
  \omega_1 = \frac{1}{6\xi}\frac{3\!-\!2\epsilon}{2\!-\!\epsilon}\epsilon
\,,\qquad \omega_2 = 1- \frac{1}{6\xi}\frac{3\!-\!2\epsilon}{2\!-\!\epsilon}\epsilon
\,.
\label{omega 1}
\end{equation}
In this case the amplitude of the stochastic force~(\ref{stochastic force:accelerated:3}) becomes cutoff independent,
\begin{equation}
 A \;\stackrel{\nu\rightarrow 3/2}{\longrightarrow }\; A_0=\frac{(1\!-\!\epsilon)^3 }{4\pi^2}
    \,,
\label{Appendix:stochastic force:accelerated:3}
\end{equation}
and the late-time limit of Eq.~(\ref{Fokker-Planck equation:2}) becomes,
\begin{equation}
 \partial_{\tilde\varphi}^2\tilde\rho_\infty
   +\frac{2}{A_0}\frac{1}{3\!-\!\epsilon}\partial_{\tilde\varphi}
                    \Big[\tilde\rho_\infty\Big(\frac{\lambda}{6}\tilde\varphi^3+6\omega_2\xi(2\!-\!\epsilon)\tilde\varphi
                                          - (3\!-\!\epsilon)\epsilon\tilde\varphi
                              \Big)
                    \Big] =0
   \,.
\label{Appendix:Fokker-Planck equation:2}
\end{equation}
This equation governs the late-time PDF of the scalar field in stochastic inflation.
As in~Eq.~(\ref{Fokker-Planck equation:2b})  above, one can introduce a rescaled effective potential,
\begin{equation}
  \bar v_{\rm eff} = \frac{2}{(3\!-\!\epsilon)A_0}\left[\tilde V(\tilde\varphi)
                             +\frac{3(2\xi\!-\!\epsilon)(2\!-\!\epsilon)}{2}\tilde\varphi^2\right]
\label{Appendix:Fokker-Planck equation:2b}
\end{equation}
and the equation of motion for $\bar v(\tilde\varphi)=-\ln[\tilde \rho_\infty(\tilde\varphi)/\tilde\rho_0]$ is,
\begin{equation}
  -\bar v''+\bar v_{\rm eff}''+[\bar v'-\bar v_{\rm eff}']\bar v' = 0
\,.
\label{Appendix:Fokker-Planck equation:2c}
\end{equation}
\vskip -0.15cm\noindent
Upon comparing with~(\ref{Fokker-Planck equation:2c}--\ref{Fokker-Planck equation:2d}), we see that the correct solution is,  
\begin{equation}
  \bar v(\tilde\varphi)=\bar v_{\rm eff}(\tilde\varphi)
\,.
\label{Appendix:Fokker-Planck equation:2d}
\end{equation}
\vskip -0.15cm\noindent
From this result we conclude that the cutoff independent solution for the field PDF inherits the tree mass term minus a mass term 
that is proportional to $\epsilon$, such that the effective mass-squared is positive (negative) when $2\xi\! -\! \epsilon$ is 
positive (negative). We emphasize again that the solution~(\ref{Appendix:Fokker-Planck equation:2d}) is fully cutoff and scale independent.

While there is no good reason to think that such a modification of the theory should correctly reproduce early time 
evolution of the field correlators, the late-time correlators may be correctly reproduced.
In order to fully understand the implications of the result~(\ref{Appendix:Fokker-Planck equation:2d}),
one would have to perform a suitable RG analysis.

\end{document}